\documentclass[12pt,a4paper]{article}
\usepackage[dvips]{graphicx}       
\usepackage{amssymb}        

\renewcommand{\baselinestretch}{1.2}

\newcommand{\atamasage}{\hspace*{1.3em}}
\newcommand{\atamaage}{\hspace*{-1.3em}}
\newcommand{\bb}[1]{$#1$\hspace{-1.6ex}%
{\tiny$^{^{(}}$}\hspace{-0.3ex}$^{-}$\hspace{-0.2ex}{\tiny$^{^{)}}$}}
\newcommand{\bbb}[1]{$#1$\hspace{-1.5ex}%
{\tiny$^{^{^{(}}}$}\hspace{-0.2ex}$^{^-}$\hspace{-0.1ex}{\tiny$^{^{^{)}}}$}}
\begin{document} 
\begin{titlepage}
   \vspace*{0cm}
  {\center
                \bf\Huge{GRAPE\,--\,Dilepton}          \vspace{-0.6cm} \\
                    {\large (Version\,1.1)}                \vspace{-0.1cm} \\
         \it\Large{A Generator for Dilepton Production   \vspace{-0.1cm} \\
                   in $ep$ Collisions} \vspace{-0.1cm} \\
  }


\vspace{0.1cm}
{\center \Large Tetsuo Abe \par}
\vspace{-0.2cm}
{\center \it Department of Physics, University of Tokyo \\
              7-3-1 Hongo, Bunkyo-ku, Tokyo 113-8654, Japan \par}

\vspace{0.8cm}

 {\center\large \today \par}

\vspace{1cm}
\begin{quotation}
\noindent
{\bf Abstract:}
{\tt GRAPE-Dilepton} is a Monte Carlo  event generator for dilepton production
  in $ep$ collisions.
The cross-section calculation is based on the exact matrix elements
    in the electroweak theory at tree level.
The dilepton productions
   via $\gamma\gamma$, $\gamma$Z$^0$, Z$^0$Z$^0$ collisions
   and via photon internal conversion 
   are taken into account.
In addition, the effects of the Z$^0$ on/off-shell production are also included.
The relevant Feynman amplitudes are generated by
        the automatic calculation system {\tt GRACE}.
The calculation of the proton vertex covers the whole kinematical region.
  This generator has an interface to {\tt PYTHIA} and {\tt SOPHIA}
        to obtain complete hadronic final states.
 \end{quotation}


\end{titlepage}

\pagestyle{empty}
  \clearpage
\section*{PROGRAM SUMMARY}

{\it Title of program}: {\tt GRAPE-Dilepton}\,(v1.1) \vspace{-0.15cm}\\

\atamaage
{\it Program obtainable from}:
   CPC Program Library, Queen's University of Belfast, N. Ireland
   and from
   {\tt http://www-zeus.desy.de/$^{\sim}$abe/grape/}. \vspace{-0.15cm}\\

\atamaage
{\it Operating system under which the program has been tested}:
   UNIX  \vspace{-0.15cm}\\

\atamaage
{\it Programming language used}: Fortran77 \vspace{-0.15cm}\\

\atamaage
{\it Memory required to execute with typical data}:
   7 Mwords for integrations,
   9 Mwords for event generations \vspace{-0.15cm}\\

\atamaage
{\it Keywords}:
  dilepton,
  lepton-pair, 
  ep collision,
  Bethe-Heitler,
  Z boson,
  dipole form factor,
  hadron tensor,
  lepton tensor,
  structure function,
  parton density,
  GRACE
   \vspace{-0.15cm}\\

\atamaage
{\it Nature of physical problem}:
 A precise estimation of the cross section of the electroweak
   dilepton production in $ep$ collisions
   is required in various physics analyses,
   where 8$\sim$48 Feynman diagrams can contribute. \vspace{-0.15cm}\\

\atamaage
{\it Method of solution}: 
 The automatic calculation system {\tt GRACE} is used
   to obtain all of the relevant helicity amplitudes.
 The phase space is divided into the 3 regions
   according to the kinematics at the proton vertex,
   and the 3 different calculation methods are applied.
 The radiative corrections are included
   using the structure function and the parton shower methods.
  \vspace{-0.15cm}\\

\atamaage
{\it Restrictions on the complexity of the problem}:
Higgs, the proton-Z$^0$ coupling and
  lepton pair production through photon radiation from the proton
  are not included.
The contribution from the resolved photon, {\it i.e.}
  Drell-Yan process in $ep$ collisions is not included. \vspace{-0.15cm}\\

\atamaage
{\it Typical running time}:
  {\it 1 hour} for a cross-section integration
  and
  {\it 1 msec} per 1 event for an event generation \vspace{-0.15cm}\\

  \clearpage
\pagestyle{plain}

\pagenumbering{roman}
\setcounter{page}{1}

\pagenumbering{arabic}
  \section{Introduction}
\atamasage
In the study of electron/positron-proton ($ep$) collisions,
  a precise estimation of
  the dilepton\footnote{The word {\it dilepton} represents
    di-electron(di-$e$), di-muon(di-$\mu$) and di-tau(di-$\tau$) in this paper.}
  production cross-sections in the electroweak (EW) interaction
  is important
    since it could become a significant background
      for various physics analyses such as, for example,
      exclusive $J/\psi$ or $\Upsilon$ production
      and
      new physics searches.
So far only the generator {\tt LPAIR} \cite{LPAIR} has been used in experimental
      analyses to estimate the dilepton background \cite{Dirk}.
    The calculation of {\tt LPAIR} is based on the diagrams of
    the photon-photon collision process \cite{GAMGAM},
    so-called {\it two-photon Bethe-Heitler} (2-$\gamma$ BH),
    corresponding to the diagrams of 
      Fig.\,\ref{diagrams_GRAPE_ela}-(a) or
      Fig.\,\ref{diagrams_GRAPE_dis}-(a)
    with the photon contribution only.
This process is dominant in most of the phase space.
It is, however, expected that 
  in the region of low invariant masses of the dilepton system,
  QED-Compton type (CO) diagrams ({\it i.e.}~photon internal conversion process)
    as seen in
    Fig.\,\ref{diagrams_GRAPE_ela}-(b) and Fig.\,\ref{diagrams_GRAPE_dis}-(b)
  become dominant.
    In the high mass region, there is an additional interesting process, 
    {\it i.e.}~Z$^{0}$ production,
  which is implemented
      in the MC event generator {\tt EPVEC} \cite{EPVEC}.
{\tt EPVEC}, however, does not include
  2-$\gamma$ BH nor CO diagrams.
In the di-$e$ channel,
  interference effects of the final state $e^-e^-$ or $e^+e^+$ 
  should also be taken into account,
  which
    are     included neither in {\tt LPAIR} nor in {\tt EPVEC}.

In this paper, a new MC event generator {\tt GRAPE-Dilepton}
  for the dilepton production in $ep$ collisions is presented.
The FORTRAN code to calculate the Feynman amplitudes
  is generated by {\tt GRACE} \cite{GRACE}
  which is an automatic calculation system.
          {\tt GRACE} has been used mainly for $e^+e^-$ interactions so far.
This is the first time for {\tt GRACE} to be applied to the case where
   there is a composite particle ({\it i.e.}~proton) in the initial state.
{\tt GRAPE} stands for a
  {\it \underline{GRA}ce-based generator for
       \underline{P}roton-\underline{E}lectron collisions}.

This generator has the following features.   \vspace{-0.3cm}
\begin{itemize}
  \item The cross-section calculation is based on the exact matrix elements
           in the electroweak theory at tree level.
        Not only 2-$\gamma$ BH but also the dilepton productions
          via $\gamma$Z$^0$  and Z$^0$Z$^0$ collisions
          are taken into account.
        CO and Z$^0$ on/off-shell production
          are also included.
                               Interference effects of the final state
           $e^{\pm}e^{\pm}$ are taken into account in the di-$e$ channel.
        It is possible to select any sub-set of diagrams in the calculation.
     \vspace{-0.2cm}
  \item All fermion masses are kept    non-zero
           both in the matrix elements and in the kinematics, which makes it
           possible to use this program with arbitrary
             small scattering angles of $e^{\pm}$  and/or
             small invariant masses of dilepton
           down to the kinematical limits.
     \vspace{-0.2cm}
  \item  The calculation of the proton vertex covers
            the whole kinematical region
            by dividing it into 3 categories of
            elastic, quasi-elastic and DIS (Deep Inelastic $eq$ Scattering)
            processes,
            as described in the next section in detail.
     \vspace{-0.2cm}
  \item  Both of
          Initial State Radiation (ISR) and Final State Radiation (FSR)
            can be included.
     \vspace{-0.2cm}
\end{itemize}

  \section{Physics aspects}
\atamasage
This generator simulates
  the $ep$ interaction:\,\,$e^{\pm}_{(in)}\,p_{_{(in)}} \hspace{-0.1cm}
                     \rightarrow e^{\pm} l^+ l^- X$
where $e^{\pm}_{(in)}$ and $p_{_{(in)}}$ indicate the
  electron/positron and the proton in the initial state respectively,
$e^{\pm}$ and $l^+ l^- $ are
  the scattered electron/positron and the produced dilepton respectively.
The relevant processes are classified into 3 categories using
  the negative momentum transfer squared at the proton vertex\,($Q_p^2$) and
  the invariant mass of the hadronic system\,($M_{had}$);
\begin{equation}
   \hspace*{-0.85cm}
   Q_{p}^2 \stackrel{\rm def}{\equiv} - \left\{
          p_{_{e^{\pm}(in)}}
       - (p_{_{e^{\pm}}} + p_{_{l^+}} + p_{_{l^-}})
                  \right\}^2,
   \label{Q2_def}
\end{equation}
\begin{equation}
   M_{had}^2 \stackrel{\rm def}{\equiv} \left\{
       (p_{_{e^{\pm}(in)}} + p_{_{p(in)}})
     - (p_{_{e^{\pm}}} + p_{_{l^+}} + p_{_{l^-}})
                \right\}^2,
   \label{Mh_def}
\end{equation}
  where $p_{_{e^{\pm}(in)}}$ and $p_{_{p(in)}}$ are the 4-momenta of
  the incoming lepton and the proton after ISR, respectively.
$p_{_{e^{\pm}}}$ and $p_{_{l^{\pm}}}$ are those of the scattered
  lepton and the produced leptons before FSR, respectively.
The 3 categories are
\vspace{-0.2cm}
\begin{itemize}
  \item  $M_{had}=M_p$\,({\it elastic}),
     \vspace{-0.2cm}
  \item  $Q_{p}^2<Q_{min}^2$ OR\,
         $M_p+M_{\pi^{0}}<M_{had}<M_{cut}$\,({\it quasi-elastic}),
     \vspace{-0.2cm}
  \item  $Q_{p}^2>Q_{min}^2$ AND $M_{had}>M_{cut}$\,({\it DIS}),
     \vspace{-0.2cm}
\end{itemize}
    where $M_p$ and $M_{\pi^{0}}$ are the masses of the proton and the neutral pion,
    respectively.
$Q_{min}$ is set to around 1\,GeV depending on the Parton Density Function (PDF)
  used in the DIS process.
The recommended value for $M_{cut}$ is 5\,GeV.

For the elastic process, the diagrams in Fig.\,\ref{diagrams_GRAPE_ela}
  are calculated with the following dipole form factor
   for the proton-proton-photon              vertex
    ($\Gamma^{\mu}_{pp\gamma}$) with the on-shell proton.
The general form of the elastic proton vertex can be written as
\begin{equation}
  \Gamma^{\mu}_{pp\gamma}
  {\small =e_p\left( F_1(Q_{p}^2)\gamma^\mu  +
          \frac{\kappa_p}{2M_p}F_2(Q_{p}^2)
          \,i\sigma^{\mu\nu}q_{\nu}\right) }
  \label{ppg_vertex_general}
\end{equation}
where  $e_p$ indicates the electric charge of the proton,
  $q$ is the 4-momentum transfer at the proton vertex ($q^2=-Q_{p}^2$),
       $F_1(Q_{p}^2)$ and $F_2(Q_{p}^2)$ are the 2 independent form factors,
         and 
           $\kappa_p$ is the anomalous magnetic moment of the proton
       (see, for example, \cite{Q_and_L}.).
The electric and magnetic form factors
    $G_E^p(Q_{p}^2)$ and $G_M^p(Q_{p}^2)$, respectively
  are defined as follows,
\begin{equation}
  \left(
      \begin{array}{@{\,}c@{\,}}
        G_E^p(Q_{p}^2) \\
        G_M^p(Q_{p}^2)
      \end{array}
         \right) =    
      \left(
      \begin{array}{@{\,}ccc@{\,}ll}
        F_1(Q_{p}^2)&-&\frac{\kappa_p Q_{p}^2}{4M_p^2} F_2(Q_{p}^2) \\
        F_1(Q_{p}^2)&+&\,\,\,\kappa_p\,\,\, F_2(Q_{p}^2)
      \end{array}
         \right).
  \label{Ge_Gm_deg}
\end{equation}
    Using the Gordon decomposition and the scaling law of the form factor,
\begin{equation}
  G_E^p(Q_{p}^2) = G_M^p(Q_{p}^2) / |\mu_p|,
  \label{Ge_Gm_scaling}
\end{equation}
the following formula which is used in this program is obtained,
\begin{equation}
  \Gamma^{\mu}_{pp\gamma}
  =e_p\left(\mu_p G_E^p(Q_{p}^2)\gamma^{\mu} -
      \frac{ (p^{\mu}_{p(in)}\hspace{-0.2cm} + p^{\mu}_{p(out)})}{2M_p}
      \frac{\kappa_p}{1+\frac{Q_{p}^2}{4M_p^2}} G_E^p(Q_{p}^2)  \right)
  \label{ppg_vertex_used}
\end{equation}
  where $\mu_p = (1 +\kappa_p)\mu_B$, $\mu_B$ is the Bohr magneton,
  and $p_{p(out)}$ indicates the 4-momentum of the scattered proton.
   $G_E^p(Q_{p}^2)$ is calculated according to the formula of the dipole fit,
\begin{equation}
  G_E^p(Q_{p}^2) = \left(1+\frac{Q_{p}^2}{0.71\,{\rm GeV}^2}\right)^{-2}.
  \label{dipole_fit}
\end{equation}

The only difference between the elastic and the quasi-elastic processes is
  the treatment of the proton vertex and
  the simulation of the hadronic final state.
    The quasi-elastic proton vertex can be described using the hadron tensor
  in the following form assuming parity and
  current conservation (for example, see \cite{Q_and_L}.),
\begin{eqnarray}
   W^{\mu\nu}
     = & & W_1\left( -g^{\mu\nu}+\frac{q^{\mu}q^{\nu}}{q^2} \right)
                                                            \nonumber \\
       &+& W_2\frac{1}{M_p^2}
           \left( p_{_{p(in)}}^{\mu}-\frac{p_{_{p(in)}}\cdot q}{q^2}q^{\mu} \right) 
           \left( p_{_{p(in)}}^{\nu}-\frac{p_{_{p(in)}}\cdot q}{q^2}q^{\nu} \right).
   \label{hadron_tensor}
\end{eqnarray}
  $W_1(Q_p^2,M_{had})$ and $W_2(Q_p^2,M_{had})$ are
  the electromagnetic proton structure functions.
The hadron tensor is contracted with
     the lepton tensor $L^{\mu\nu}$ numerically to obtain the cross section,
\begin{equation}
   d\sigma \sim L_{\mu\nu} W^{\mu\nu}
   \label{x_sec_qela}.
\end{equation}
In this version, 
$W_1$ and $W_2$ are parameterized with Brasse et al.\,\cite{BRASSE}
  for $M_{had}<$\,2\,GeV (the proton resonance region),
  and with ALLM97 \cite{ALLM97} for $M_{had}>$\,2\,GeV.
 These two parameterizations are based on fits to the experimental data
   on the measurement of the total $\gamma^*p$ cross-sections.
The exclusive hadronic final state is generated
   using the MC event generator {\tt SOPHIA} \cite{SOPHIA}
   in the event generation step.

For the DIS process with the Quark Parton Model,
  the diagrams in Fig.\,\ref{diagrams_GRAPE_dis} are calculated.
{\tt PDFLIB} \cite{PDFLIB} is linked to obtain parton densities
  with $Q_p^2$ as a QCD scale.
The simulation of the proton remnant and the hadronization are
  performed by {\tt PYTHIA} \cite{PYTHIA}.
It should be noted that the lowest order calculation in this process is valid only
   for the region of
\begin{equation}
             u \stackrel{\rm def}{\equiv}
             \bigl|\bigl\{  p_{_{q(in)}} - 
                            (p_{_{l^+}} + p_{_{l^-}})  \bigr\}^2\bigr|
             \,\,\,\gtrsim\,\,\, 25\,{\rm GeV}^2,
\end{equation}
    where $p_{_{q(in)}}$ is the 4-momentum of the incoming quark.
The value of $u$ corresponds to the virtuality of the $u$-channel quark
  in the diagrams in Fig.\,\ref{diagrams_GRAPE_dis}-(b),(c).
When it is nearly or smaller than 25 GeV$^2$, 
  the lowest order calculation is not correct as explained in \cite{EPVEC}
  since QCD corrections become large.
In this case, the dilepton production should be treated as Drell-Yan process
  between the proton and the resolved photon from the beam lepton,
  which is not implemented in this program.
The cut: $u>25$\,GeV$^2$ is explicitly applied in this program
  if the diagrams other than BH are included.

The effect of ISR is included in the cross-section calculation using
  the structure function method described in \cite{ISR_SF},
  where the momentum transfer squared on the beam lepton,
  {\it i.e.} $\bigl\{ p_{_{e^{\pm}(in)}} - p_{_{e^{\pm}}}\bigr\}^2$
  is used as a QED scale.
When ISR turns on, the correction for the photon self energy,
  {\it i.e.} the vacuum polarization,
  is included according to the parameterization in \cite{QEDVAC}
  by modifying photon propagators.
FSR is performed by {\tt PYTHIA} using the parton shower method
  when the event is generated.

\begin{figure}[ptb]
  \begin{center}
     \includegraphics[scale=1.0,clip]{pictures/diagrams_GRAPE_ela}
  \end{center}
  \caption[Feynman diagrams included in the (quasi-)elastic process.]
          {Feynman diagrams included in the (quasi-)elastic process.
           $e$=$\left\{ e^+,e^- \right\}$,
           l$^{\pm}$=$\left\{ e^{\pm}, \mu^{\pm}, \tau^{\pm} \right\}$.
           N means a (dissociated) proton or a nucleon resonance.
          }
  \label{diagrams_GRAPE_ela}
\end{figure}

\begin{figure}[ptb]
  \begin{center}
     \includegraphics[scale=0.8,clip]{pictures/diagrams_GRAPE_dis}
  \end{center}
  \caption[Feynman diagrams included in the DIS process.]
          {Feynman diagrams included in the DIS process.
           $e$=$\left\{ e^+,e^- \right\}$,
           l$^{\pm}$=$\left\{ e^{\pm}, \mu^{\pm}, \tau^{\pm} \right\}$ and
           $q$=$\left\{\right.${\bb u},{\bbb d},{\bb s},{\bb c},{\bbb b},{\bbb t}%
$\left.\right\}$.
          }
    \label{diagrams_GRAPE_dis}
\end{figure}

  \section{Program structure}
\begin{figure}[ptb]
  \begin{center}
     \includegraphics[scale=0.73,clip]{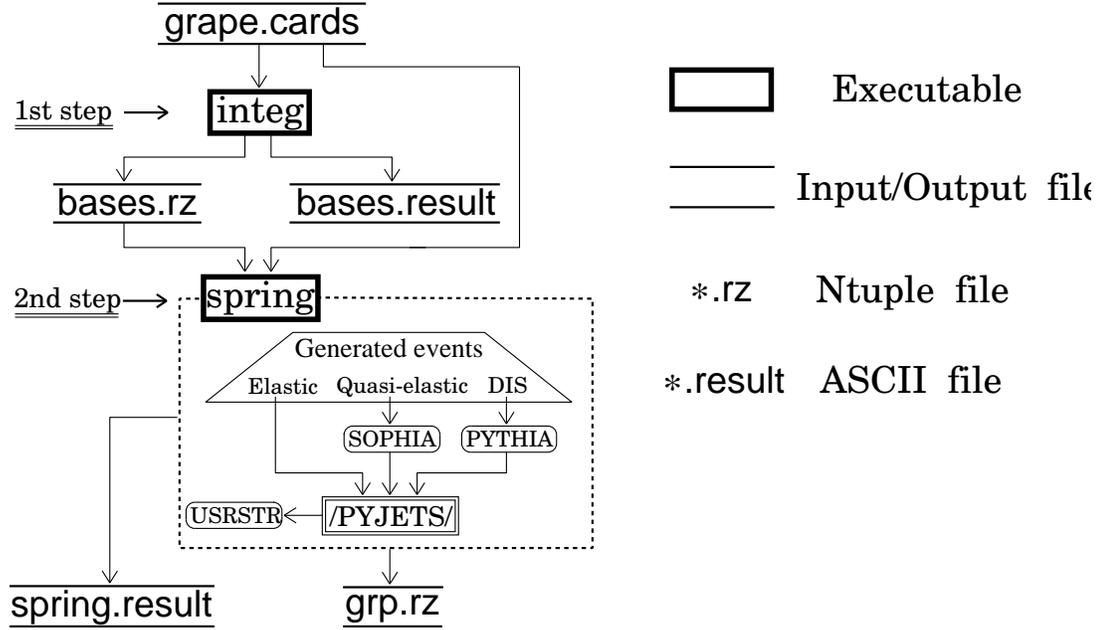}
  \end{center}
  \caption[Flowchart for the program structure]
          {Flowchart for the program structure}
  \label{flow}
\end{figure}

\atamasage
Physics events are generated with the 2 steps;
  the MC integration step by the executable:\,{\tt integ} and
  the event generation step by the executable:\,{\tt spring},
  as illustrated in Fig.\,\ref{flow}.
In both steps, the program is controlled
  by an ASCII file:\,{\tt grape.cards}.
The file is read by the executables
   with     help of {\tt FFREAD} \cite{FFREAD}.
The contents of {\tt grape.cards} are explained in the next section.

In the integration step by the executable:\,{\tt integ},
  an effective total cross-section (in unit of pb) and
  probability distributions are calculated by {\tt BASES} \cite{BS}.
The results are stored in a file:\,{\tt bases.rz}
  which has the Ntuple format provided by the {\tt HBOOK} package \cite{HBOOK}.
At the same time, 
  the information related to
  the convergency status of the integration is output into
  an ASCII file:\,{\tt bases.result}.

In the event generation step by the executable:\,{\tt spring}, 
  unweighted events are generated.
This is done by an routine:\,{\tt SPRING} \cite{BS}
  according to the probability distributions in {\tt bases.rz}.
The results of the event generation are stored
  in the {\tt PYTHIA} common block {\tt /PYJETS/}.
After filling {\tt /PYJETS/}, {\tt spring} calls a routine:\,{\tt USRSTR}
  in which user specific procedures are put.
Its template is found in the appendix.
The event information in {\tt /PYJETS/} is also available
  in a Ntuple file:\,{\tt grp.rz}.

The calculated cross-section is found in {\tt bases.result} or
  at the end of the standard output from {\tt spring}.
The status of the event generation is output into
  an ASCII file:\,{\tt spring.result}.
Looking at the file,
  users should find a reasonable agreement between
  generated distributions by {\tt spring} and calculated ones by {\tt integ}.
The procedure to make the executables
  is described in the {\tt README} file.

  \section{Input data cards}
\newcommand{\at}{\hspace*{1.2em}}
\newcommand{\atat}{\hspace*{2.4em}}
\atamasage
The input data in {\tt grape.cards} are explained in this section.
All of the items are optional and are set to default values if not specified.
Default values are written in the brackets starting with {\tt D=}.
The items are not explicitly displayed
  in case that they are the only one for their cards.

\vspace{0.1cm}
\begin{list}{$\bullet$}{%
\setlength{\itemsep}{.0pt}%
\setlength{\parsep}{.0pt}%
\setlength{\leftmargin}{0.4cm}%
}
\item {\bf KFLBEAM} \\
  KF code of the lepton beam
  ({\tt INTEGER,\,D=-11}); 11:electron, -11:positron.
\item {\bf EPOL}$~~~P~~~\theta~~~\phi$ \\
  Polarization of the lepton beam ({\tt REAL}); \\
\at    $P$ = degree of the polarization in the range $[-1,1]$ ({\tt D=0.}), \\
\at\hspace*{0.12cm}$\theta$ = polar angle of the polarization vector in degree
    ({\tt D=0.}), \\
\at\hspace*{0.08cm}$\phi$ = azimuthal angle of the polarization vector in degree
    ({\tt D=0.}). \\
 The positive direction of the z-axis on the polarization vector
 is in the direction of the lepton beam.
\item {\bf EBEAM} \\
 Lepton beam momentum in {\bf MeV/c}
    ({\tt REAL,\,D=27520.}).
\item {\bf PBEAM}  \\
 Proton beam momentum in {\bf MeV/c}
    ({\tt REAL,\,D=820000.}).
\item {\bf PROCESS}  \\
 Process type of the proton vertex
    ({\tt INTEGER,\,D=1}); \\
\at   1:elastic, 2:quasi-elastic, 3:DIS.
\item {\bf LPAIR}  \\
 Dilepton channel
    ({\tt INTEGER,\,D=2}); 1:di-$e$, 2:di-$\mu$, 3:di-$\tau$.
%
\item {\bf ISR}  \\
 Initial state radiation flag for the beam lepton
    ({\tt INTEGER,\,D=1}); 0:off, 1:on.
%
\item {\bf QFLV}  \\
 Scattered quark in the DIS process ({\tt INTEGER,\,D=1}); \\
\at  1:$u$, 2:$\bar{u}$, 3:$d$, 4:$\bar{d}$, 5:$s$, 6:$\bar{s}$, 7:$c$,
      8:$\bar{c}$, 9:$b$, 10:$\bar{b}$, 11:$t$, 12:$\bar{t}$.
%
\item {\bf MERGE}  \\
 Merging mode in the DIS process ({\tt INTEGER,\,D=0}); 0:off. \\
       In some cases,
       contributions from different quarks can be included
       in the cross-section calculation adding the parton densities
       if the mass difference is negligible.
       The possible combinations of  {\bf QFLV} and {\bf MERGE}
         are written in Table\,\ref{merge}.
       The mass of the quark specified with {\bf QFLV} is used in the amplitude
       and the kinematics calculations.
\newcommand{\mergeAA}{$u+\bar{u}+d+\bar{d}$}
\newcommand{\mergeAB}{\mergeAA$\,+\,s+\bar{s}$}
\newcommand{\mergeAC}{\mergeAB $\,+\,c+\bar{c}$}
\newcommand{\mergeAD}{\mergeAC $\,+\,b+\bar{b}$}

\newcommand{\mergeBA}{$u+c$}
\newcommand{\mergeBB}{$\bar{u}+\bar{c}$}

\newcommand{\mergeCA}{$d+s$}
\newcommand{\mergeCB}{$\bar{d}+\bar{s}$}

\newcommand{\mergeDA}{$d+s+b$}
\newcommand{\mergeDB}{$\bar{d}+\bar{s}+\bar{b}$}

\newcommand{\qflv}{{\bf QFLV}}
\newcommand{\merge}{{\bf MERGE}}

\newcommand{\YES}{\bf{Yes}}
\newcommand{\NO}{No}

\begin{table}
  \begin{center}
    {\footnotesize
    \begin{tabular}{|c|l|l||c|c|}
      \hline
       \qflv  &  \merge      &   Quarks   &     BH     & QED/EW/CO/Z$^0$\\
      \hline
      \hline
        1     &  1234        &  \mergeAA  &  \YES  &   \NO         \\
      \hline
        1     &  123456      &  \mergeAB  &  \YES  &   \NO        \\
      \hline
        1     &  12345678    &  \mergeAC  &  \YES  &   \NO     \\
      \hline
        1     &  1234567890  &  \mergeAD  &  \YES  &   \NO     \\
      \hline
        1     &  17          &  \mergeBA  &  \YES  &  \YES   \\
      \hline
        2     &  28          &  \mergeBB  &  \YES  &  \YES \\
      \hline
        3     &  35          &  \mergeCA  &  \YES  &  \YES    \\
      \hline
        4     &  46          &  \mergeCB  &  \YES  &  \YES    \\
      \hline
        3     &  359         &  \mergeDA  &  \YES  &  \YES  \\
      \hline
        4     &  460         &  \mergeDB  &  \YES  &  \YES  \\
      \hline
    \end{tabular}
    }
  \end{center}
  \caption[{\bf MERGE}]{Possible combinations of {\bf QFLV} and {\bf MERGE}}
  \label{merge}   
\end{table}
%
\item {\bf NGROUP} \\
 Author group described in the {\tt PDFLIB} manual
      ({\tt INTEGER,\,D=5}).
\item {\bf NSET} \\
 PDF set described in the {\tt PDFLIB} manual
      ({\tt INTEGER,\,D=5}).\\
 The default is GRV94(LO).
\item {\bf GRASEL} \\
 Feynman diagram selection ({\tt INTEGER,\,D=3}); \\
\at   1\,:\,2-$\gamma$ Bethe-Heitler
     (without $e^{\pm}e^{\pm}$ interference in case of di-$e$), \\
\at   2\,:\,2-$\gamma$ Bethe-Heitler
     (including $e^{\pm}e^{\pm}$ interference in case of di-$e$), \\
\at   3\,:\,QED diagrams\,({\it i.e.} all the diagrams
                                except for the Z$^0$ contribution), \\
\at   4\,:\,EW diagrams\,({\it i.e.} all the diagrams), \\
\at  13\,:\,QED-Compton type diagrams only, \\
\at  14\,:\,Z$^0$ production diagrams only. \\
In case of di-$\mu,\tau$, the first and the second selections give
 the same result.
%
%
%
\item {\bf ITMX1} \\
 Number of iterations in the grid optimization step of {\tt BASES}.
       ({\tt INTEGER,\,D=4}). \\
 This should be larger than 2.
\item {\bf ITMX2} \\
 Number of iterations in the integration step of {\tt BASES}
       ({\tt INTEGER,\,D=10}). \\
 This should be larger than 5.
\item {\bf NCALL} \\
 Number of sampling points in each iteration of {\tt BASES}
       ({\tt INTEGER,\,D=1000000}). \\
 This should be large
   so that any {\it accuracy} of each iteration
   in the integration step of {\tt BASES}
   is better than 0.5\,\%.
%
\item {\bf NGEN} \\
 Number of events to be generated by {\tt spring}
       ({\tt INTEGER,\,D=100}).
\item {\bf NMOD}$~~~N_{mod}$ \\
 Printing a message per $N_{mod}$ events in the event generation
       ({\tt INTEGER,\,D=1000}).
%
\item {\bf PSISR} \\
 Switch for the initial state parton shower by {\tt PYTHIA}
       ({\tt INTEGER,\,D=1}); 0:off, 1:on. \\
 This has an effect only on event generations of the DIS process.
 No effect on elastic and quasi-elastic events.
 This item is copied to {\tt MSTP(61)} in the {\tt PYTHIA} common block
        {\tt /PYPARS/}.
%
\item {\bf PSFSR} \\
 Switch for the final state parton shower by {\tt PYTHIA}
       ({\tt INTEGER,\,D=1}); 0:off, 1:on. \\
 This item is copied to {\tt MSTP(71)} in the {\tt PYTHIA} common block
        {\tt /PYPARS/}.
%
\item {\bf PSBRA} \\
 Parton shower branchings in {\tt PYTHIA}
       ({\tt INTEGER,\,D=2}); \\
\at  1:QCD, 2:QCD+QED. \\
 This item is copied to {\tt MSTJ(41)} in the {\tt PYTHIA} common block
        {\tt /PYDAT1/}.
\item {\bf PSSUP} \\
 Suppression of the {\tt PYTHIA} parton shower
      ({\tt INTEGER,\,D=0}); 0:off, $>$$=$1:on.\\
 This item is copied to {\tt MSTJ(40)} in the {\tt PYTHIA} common block
        {\tt /PYDAT1/}.
\item {\bf PYDECAY} \\
 Switch for fragmentation and decay in {\tt PYTHIA} ({\tt INTEGER,\,D=1});
       0:off, 1:on. \\
 No effect on elastic and quasi-elastic events. \\
 This item is copied to {\tt MSTP(111)} in the {\tt PYTHIA} common block
        {\tt /PYPARS/}.
\item {\bf PRIPT} \\
 Primordial $k_t$ distribution in the proton ({\tt INTEGER,\,D=1});\\
\at  0:off, 1:gaussian, 2:exponential. \\
 No effect on elastic and quasi-elastic events. 
 This item is copied to {\tt MSTP(91)} in the {\tt PYTHIA} common block
        {\tt /PYPARS/}.
\item {\bf PYLIST} \\
 Printing the contents of {\tt /PYJETS/} ({\tt LOGICAL,\,D=TRUE}).
\item {\bf NLIST} \\
 Number of events whose {\tt /PYJETS/} is printed out
       ({\tt INTEGER,\,D=10}).
\item {\bf NTPYT} \\
 Output of generated events into a Ntuple file:\,{\tt grp.rz} \\
 from the {\tt PYTHIA} common block {\tt /PYJETS/}
    ({\tt LOGICAL,\,D=FALSE}).  \\
 The meanings of the Ntuple variables are in the following. \\
\atat    {\tt npy} :
\atat      Number of particles ({\tt integer})\\
\atat    {\tt px(1:npy),\,py(1:npy),\,pz(1:npy)} :
             x,y,z-component of momentum \\
\atat        \hspace*{6.8cm} in GeV/c ({\tt real*4})\\
\atat    {\tt pe(1:npy)} : Energy in GeV ({\tt real*4})\\
\atat    {\tt pm(1:npy)} : Mass in GeV ({\tt real*4})\\
\atat    {\tt kf(1:npy)} : KF code ({\tt integer})\\
\atat    {\tt sta(1:npy)} : Status code ({\tt integer})\\
\atat    {\tt mot(1:npy)} : Line number of the mother particle ({\tt integer})
\item {\bf Q2RNGME}$~~~Min~~~Max$ \\
   Range for the negative momentum transfer squared at the electron vertex
   $Q^2_e$ \underline{without} ISR ({\tt REAL}),
   {\it i.e.} $Q^2_e = - \bigl\{ p_{_{e^{\pm}(in)}} - p_{_{e^{\pm}}}\bigr\}^2$
   where $p_{_{e^{\pm}(in)}}$ is a 4-momentum of the incoming lepton
   \underline{after} ISR. \\
\at\hspace*{0.63mm} $Min$ = the minimum in GeV$^2$ ({\tt D=0.}), \\
\at\hspace*{0.00mm} $Max$ = the maximum in GeV$^2$ ({\tt D=1.E20}). \\
 In case of di-$e$ with $e^{\pm}e^{\pm}$ interference,
  smaller one of the two $Q^2_e$ values is used.
\item {\bf Q2RNGOB}$~~~Min~~~Max$ \\
   Range for the negative momentum transfer squared at the electron vertex
   $Q^2_e$ \underline{including} ISR ({\tt REAL}),
   {\it i.e.} $Q^2_e = - \bigl\{ p_{_{e^{\pm}(in)}} - p_{_{e^{\pm}}}\bigr\}^2$
   where $p_{_{e^{\pm}(in)}}$ is a 4-momentum of the incoming lepton
   \underline{before} ISR. \\
\at\hspace*{0.63mm} $Min$ = the minimum in GeV$^2$ ({\tt D=0.}), \\
\at\hspace*{0.00mm} $Max$ = the maximum in GeV$^2$ ({\tt D=1.E20}). \\
 In case of di-$e$ with $e^{\pm}e^{\pm}$ interference,
  smaller one of the two $Q^2_e$ values is used.
\item {\bf MHAD}$~~~Min~~~Max$ \\
 Range for the mass of the hadronic system $M_{had}$ ({\tt REAL}); \\
\at\hspace*{0.63mm} $Min$ = the minimum in GeV ({\tt D=1.08}), \\
\at\hspace*{0.00mm} $Max$ = the maximum in GeV ({\tt D=1.E20}). \\
 No effect on elastic events.
\item {\bf Q2P}$~~~Min~~~Max$ \\
 Range for the negative momentum transfer squared at the proton vertex $Q_p^2$
      ({\tt REAL}); \\
\at\hspace*{0.63mm} $Min$ = the minimum in GeV$^2$ ({\tt D=0.}), \\
\at\hspace*{0.00mm} $Max$ = the maximum in GeV$^2$ ({\tt D=1.E20}). \\
 In case of the DIS process, $Q_p^2$ is used as a QCD scale for PDF.
\item {\bf THMIN}\hspace{1.7mm}
   $~~~\theta_{min}^{(1)}
    ~~~\theta_{min}^{(2)}
    ~~~\theta_{min}^{(3)}
    ~~~\theta_{min}^{(4)}$
\item {\bf THMAX}\hspace{0.0mm}
   $~~~\theta_{max}^{(1)}
    ~~~\theta_{max}^{(2)}
    ~~~\theta_{max}^{(3)}
    ~~~\theta_{max}^{(4)}$
\item {\bf EMIN}\hspace{4.7mm}
   $~~~E_{min}^{(1)}
    ~~~E_{min}^{(2)}
    ~~~E_{min}^{(3)}
    ~~~E_{min}^{(4)}$
\item {\bf EMAX}\hspace{3.1mm}
   $~~~E_{max}^{(1)}
    ~~~E_{max}^{(2)}
    ~~~E_{max}^{(3)}
    ~~~E_{max}^{(4)}$
\item {\bf PMIN}\hspace{4.6mm}
   $~~~P_{min}^{(1)}
    ~~~P_{min}^{(2)}
    ~~~P_{min}^{(3)}
    ~~~P_{min}^{(4)}$
\item {\bf PMAX}\hspace{3.0mm}
   $~~~P_{max}^{(1)}
    ~~~P_{max}^{(2)}
    ~~~P_{max}^{(3)}
    ~~~P_{max}^{(4)}$
\item {\bf PTMIN}\hspace{1.3mm}
   $~~~Pt_{min}^{(1)}
    ~~~Pt_{min}^{(2)}
    ~~~Pt_{min}^{(3)}
    ~~~Pt_{min}^{(4)}$
\item {\bf PTMAX}\hspace{-0.3mm}
   $~~~Pt_{max}^{(1)}
    ~~~Pt_{max}^{(2)}
    ~~~Pt_{max}^{(3)}
    ~~~Pt_{max}^{(4)}$ \\
\at  $(1)$ : for scattered proton or quark, \\
\at  $(2)$ : for scattered $e^{\pm}$, \\
\at  $(3)$ : for produced $l^{\mp}$, \\
\at  $(4)$ : for produced $l^{\pm}$. \\
  The above 8 data cards are used for describing the detector cut
    in the laboratory frame ({\tt REAL}).
  {\bf Each} final state particle is required to satisfy the following,\\
    \centerline{\hspace*{0.86cm}
                $\theta_{min}^{(i)} < \theta < \theta_{max}^{(i)}$
    ~{\rm AND}~ $E_{min}^{(i)} < E < E_{max}^{(i)}$
               }
    \centerline{{\rm AND}~ 
                $P_{min}^{(i)} < P < P_{max}^{(i)}$
    ~{\rm AND}~ $Pt_{min}^{(i)} < Pt < Pt_{max}^{(i)}$
               }
  where $\theta$(degree), $E$(GeV), $P$(GeV/c) and $Pt$(GeV/c)
    indicate polar angle, energy, momentum and transverse momentum, respectively.
  The default values correspond to not applying this cut.
%
\item {\bf THPTMCT}$~~~\theta_{min}~~~\theta_{max}$
\item {\bf PTMXCT}$~~~~Pt_{min}~~~Pt_{max}$ \\
 Using the above 2 data cards,
  final state leptons are required to satisfy the following ({\tt REAL}), \\
    \centerline{\hspace*{0.86cm}
                $\theta_{min} < \theta^M < \theta_{max}$
    ~{\rm AND}~ $Pt_{min} < Pt^M < Pt_{max}$
               }
  where $Pt^M$(GeV/c) indicates the maximum transverse momentum
    among the 3 final state
                     leptons\,($e^{\pm},l^{\mp},l^{\pm}$),
    and $\theta$(degree) is the polar angle of the lepton with $Pt^M$.
  The default values correspond to not applying this cut.
%
\item {\bf MASSLL}$~~~~Min1~~~Max1~~~~~Min2~~~Max2$ \\
 Range for the mass of the produced dilepton system ({\tt REAL}); \\
\at\hspace*{0.63mm} $Min1$ = the minimum in GeV ({\tt D=0.}), \\
\at\hspace*{0.00mm} $Max1$ = the maximum in GeV ({\tt D=1.E20}). \\
 In case of di-$\mu$,\,$\tau$, $Min2$ and $Max2$ are not used.\\
 In case of di-$e$ with $e^{\pm}e^{\pm}$ interference,
      there are two masses;\,$M_{e^+e^-}^{(1)}$,$M_{e^+e^-}^{(2)}$\\
   ($M_{e^+e^-}^{(1)} < M_{e^+e^-}^{(2)}$),
      and they are required to satisfy the following,\\
    \centerline{\hspace*{0.86cm}
                $Min1 < M_{ee}^{(1)} < Max1$
    ~{\rm AND}~ $Min2 < M_{ee}^{(2)} < Max2$.
               }
\item {\bf MASSELL}$~~~Min~~~Max$ \\
 Range for the mass of the final state
      lepton system of $e^{\pm}\,l^{\mp}\,l^{\pm}$ ({\tt REAL}); \\
\at\hspace*{0.63mm} $Min$ = the minimum in GeV ({\tt D=1.}), \\
\at\hspace*{0.00mm} $Max$ = the maximum in GeV ({\tt D=1.E20}).
\item {\bf MASSQLL}$~~~Min~~~Max$ \\
 Range for the mass of the scattered quark and produced dilepton
   system of $q\,\,l^+ l^-$ ({\tt REAL}); \\
\at\hspace*{0.63mm} $Min$ = the minimum in GeV ({\tt D=5.}), \\
\at\hspace*{0.00mm} $Max$ = the maximum in GeV ({\tt D=1.E20}). \\
   This cut has an effect only on the DIS process. \\
   In case of di-$e$ with $e^{\pm}e^{\pm}$ interference,
     smaller one of the 2 values is used.
%
\item {\bf IVISI} \hspace{1.6cm} $N_{visi}$
\item {\bf THEVMIN}\hspace{1.7mm}
   $~~~\theta_{min}^{(1)}
    ~~~\theta_{min}^{(2)}
    ~~~\theta_{min}^{(3)}
    ~~~\theta_{min}^{(4)}$ 
\item {\bf THEVMAX}\hspace{0.0mm}
   $~~~\theta_{max}^{(1)}
    ~~~\theta_{max}^{(2)}
    ~~~\theta_{max}^{(3)}
    ~~~\theta_{max}^{(4)}$
\item {\bf EVMIN}\hspace{7.7mm}
   $~~~E_{min}^{(1)}
    ~~~E_{min}^{(2)}
    ~~~E_{min}^{(3)}
    ~~~E_{min}^{(4)}$
\item {\bf EVMAX}\hspace{6.1mm}
   $~~~E_{max}^{(1)}
    ~~~E_{max}^{(2)}
    ~~~E_{max}^{(3)}
    ~~~E_{max}^{(4)}$
\item {\bf PTVMIN}\hspace{4.0mm}
   $~~~Pt_{min}^{(1)}
    ~~~Pt_{min}^{(2)}
    ~~~Pt_{min}^{(3)}
    ~~~Pt_{min}^{(4)}$
\item {\bf PTVMAX}\hspace{2.3mm}
   $~~~Pt_{max}^{(1)}
    ~~~Pt_{max}^{(2)}
    ~~~Pt_{max}^{(3)}
    ~~~Pt_{max}^{(4)}$ \\
\at  $(1)$ : for scattered proton or quark, \\
\at  $(2)$ : for scattered $e^{\pm}$, \\
\at  $(3)$ : for produced $l^{\mp}$, \\
\at  $(4)$ : for produced $l^{\pm}$. \\
  The above 6 data cards are used for describing the detector cut
    in the laboratory frame ({\tt REAL} except for $N_{visi}$:{\tt INTEGER}).
  $N_{visi}$ particle(s) are(is) required to satisfy the following,\\
    \centerline{\hspace*{0.86cm}
                $\theta_{min}^{(i)} < \theta < \theta_{max}^{(i)}$
    ~{\rm AND}~ $E_{min}^{(i)} < E < E_{max}^{(i)}$
    ~{\rm AND}~ $Pt_{min}^{(i)} < Pt < Pt_{max}^{(i)}$
               }
  where $\theta$(degree), $E$(GeV) and $Pt$(GeV/c)
    indicate polar angle, energy and transverse momentum, respectively.
  As for $N_{visi}$, {\tt D=-1}, which corresponds to not applying this cut.
  The test run at the end of this paper is instructive for understanding
  this cut.
\end{list}%

  \section{Summary}
\atamasage
A new Monte Carlo generator for dilepton production
  in the framework of the electroweak theory
  was presented.
The whole kinematical region on the proton vertex is covered.
This generator can be used for quantitative and precise estimations
  of processes which
  come in addition to the two-photon Bethe-Heitler contributions.

  \section*{Acknowledgements}
\atamasage
I am very thankful to
  Prof.\,\,Y.\,Shimizu,
  Prof.\,\,S.\,Yamada and
  Prof.\,\,K.\,Tokushuku
    for their useful comments and various supports.
I shall never forget the collaboration with
  the members of the Minami-Tateya group at KEK;
  J.\,Fujimoto, T.\,Ishikawa, T.\,Kaneko, K.\,Kato, S.\,Kawabata, T.\,Kon,
  Y.\,Kurihara, H.\,Tanaka and T.\,Watanabe.
I also wish to thank the conveners of
  the working group on QED radiative effects
  in HERA Monte Carlo Workshop at DESY;
  L.\,Favart, S.\,Schlenstedt and H.\, Spiesberger for the helpful discussions.
This work was supported in part by the Japanese Ministry of Education,
  Science and Culture\,(the Monbusho) and its grant for Scientific Research
  (No.11206203).


\clearpage
\section*{Appendix A. Event store in {\tt /PYJETS/}}
\atamasage 
Each line in the {\tt PYTHIA} event store
  has the following meaning
  according to the {\tt PYTHIA} convention.

\vspace{0.3cm}
    \begin{tabular}{|c|l|}
      \hline
       Line number   &   Meaning \\
      \hline
      \hline      
           1,2       &   Beam particles (1:$p$, 2:$e^{\pm}$) \\
      \hline      
           3,4       &   Partons from the beam particles {\bf before} ISR
                           \vspace{-0.15cm} \\
                     &   (3:from $p$, 4:from $e^{\pm}$)   \\
      \hline      
           5,6       &   Partons from the beam particles {\bf after}  ISR;
                           \vspace{-0.15cm} \\
                     &   \hspace*{0.3cm}
                         the initial state in the matrix element calculation
                           \vspace{-0.15cm} \\
                     &   (5:from $p$, 6:from $e^{\pm}$)   \\
      \hline      
         7,8,9,10    &   Final state particles {\bf before} FSR;
                           \vspace{-0.15cm} \\
                     &   \hspace*{0.3cm}
                         the final state in the matrix element calculation
                           \vspace{-0.15cm} \\
                     &  (7:$p$, 8:$e^{\pm}$, 9:$l^{\mp}$, 10:$l^{\pm}$) \\
      \hline      
         11$\sim$    &   Final state particles
                           {\bf after} FSR, fragmentation and decay \\
      \hline
    \end{tabular}
\vspace{0.8cm}

\atamaage
 In case of the (quasi-)elastic process,\vspace{-0.1cm}
 \begin{itemize}
    \item  a parton from the beam proton is the proton itself,
           {\it i.e.} the 1st and the 3rd lines are the same,\vspace{-0.2cm}
    \item  the beam proton always makes no ISR, so that
           the 1st, 3rd and 5th lines are the same,\vspace{-0.2cm}
    \item  the 2nd and the 4th lines are also the same.
 \end{itemize}

\atamaage
 In case of the DIS process with both ISR and FSR,
   all of the lines have different contents in general.

 As for di-$e$ events, there are 2 identical particles in the final state.
 In {\tt GRAPE},  those 2 particles are distinguished
   in the following way;\vspace{-0.1cm}
      \begin{itemize}
        \item  in case of $e^{\pm}e^{\pm}$ interference {\bf off},
               a particle stored in the 8th line is a scattered lepton, and
               one in the 10th line
                 comes from the 2-$\gamma$ collision
                 ({\it i.e.} a produced lepton),
          \vspace{-0.2cm}
        \item  in case of $e^{\pm}e^{\pm}$ interference {\bf on},
               a lepton stored in the 8th line has smaller transverse momentum
               than that of a lepton in the 10th line.
      \end{itemize}

\clearpage
\section*{Appendix B. Routines/function related to users}

$\bullet\,${\bf Function} {\tt DRN(ISEED)} \\
provides uniform random numbers.
All other random number routines are linked to this one.
This routine is stored in the {\tt BASES} library\,({\tt libbases.a}).\\
$\bullet\,${\bf Subroutine} {\tt  DRNSET(ISEED)} \\
performs an initialization for {\tt DRN(ISEED)}.
This is also stored in {\tt libbases.a}. \\
$\bullet\,${\bf Subroutine} {\tt READ$\_$CARDS(LUN,filename)} \\
reads input data cards from {\tt grape.cards}.\\
$\bullet\,${\bf Subroutine} {\tt  SETMAS} \\
provides masses/widths of particles and the QED coupling constant. \\
$\bullet\,${\bf Subroutine} {\tt USRSTR(Ievt,Ngen)} \\
can be modified by users to access the information
  on generated events.
User initialization/termination procedures are also
  put in this routine. \\

\clearpage
\section*{Appendix C. User event storing routine: {\tt USRSTR}}

This routine is called $(N_{gen}+2)$ times
  where $N_{gen}$ is the number of generated events.
The 1st call is for the user initialization and
  the last one for the user termination phase.
The following is the template file prepared
  in the {\tt GRAPE} package\,({\tt usrstr.f}).

\vspace{0.3cm}
{\footnotesize
\renewcommand{\baselinestretch}{0.8}
\begin{verbatim}  
      subroutine USRSTR(Ievt, Ngen)
      implicit NONE
*------------ Arguments ------------
      integer  Ievt, Ngen
* Ngen : # of events to be generated
* Ievt : Counter --- < 1      ===> Initialization   phase
*                    1 - Ngen ===> Event generation phase
*                    > Ngen   ===> Termination      phase
*-----------------------------------
*---------- PYTHIA common ----------
      integer           N,NPAD, K(4000,5)
      double precision  P(4000,5), V(4000,5)
       common /PYJETS/ N,NPAD,K,P,V              !!! Event Record !!!

      integer           MINT(400)
      double precision  VINT(400)
       common /PYINT1/ MINT,VINT
* (See PYTHIA manual for details.)
*-----------------------------------
*--------- Local variables ---------
      integer     LUN1,    LUN2,    LUN3
       parameter (LUN1=41, LUN2=42, LUN3=43)
* (You can use the above logical unit numbers.)
*-----------------------------------

******** Initialization of USER Event Storing *******
      if (Ievt .LT. 1) then

      endif
*****************************************************

************* <<< USER Event Storing >>> ************
      if ((Ievt .GE. 1).and.(Ievt .LE. Ngen)) then

      endif
*****************************************************

********* Termination of USER Event Storing *********
      if (Ievt .GT. Ngen) then

      endif
*****************************************************

      return
      end
\end{verbatim}
}



\clearpage
\section*{TEST RUN INPUT AND OUTPUT}
\subsection*{An example of the input data cards}
\atamasage
 One example is presented with the following condition;
 \begin{itemize}
   \item  process\,:\,$e^+ \,q \rightarrow e^+ \,q \,\,\mu^+ \mu^-$
          ($q$=$\left\{\right.${\bb u},{\bbb d},{\bb s}$\left.\right\}$
          with GRV94(LO)\,\cite{GRV94}), \vspace{-0.2cm}
   \item  BH diagrams only, \vspace{-0.2cm}
   \item  70\,\% polarization of the $e^+$ beam
            in the direction of the proton beam, \vspace{-0.2cm}
   \item  cuts\,: (1) $\&$ (2) $\&$ (3) $\&$ (4), \vspace{-0.2cm}
      \begin{itemize}
        \def\labelitemii{(1)}
        \item  $Q^2_p > 1$\,GeV$^2$ \,\&\, $M_{had}>5$\,GeV,
          \vspace{-0.1cm}
        \def\labelitemii{(2)}
        \item  for scattered $q$, $\theta>10^{\circ}$ $\&$ P$_t >$ 15\,GeV/c,
          \vspace{-0.1cm}
        \def\labelitemii{(3)}
        \item  (invariant mass of $\mu^+\mu^- $) $>$ 4\,GeV,
          \vspace{-0.1cm}
        \def\labelitemii{(4)}
        \item  2 of the following 3 requirements for the final state
               leptons are satisfied,
          \begin{itemize}
            \def\labelitemiii{--}
            \item for $e^+$\hspace{0.124cm}:\hspace{0.39cm}5$^{\circ}$
                                    $<\theta<$ 175$^{\circ}$
                                    $\&$ P$_t>$ 5\,GeV,
            \def\labelitemiii{--}
              \vspace{-0.1cm}
            \item for $\mu^+$\,: 20$^{\circ}$ $<\theta<$ 160$^{\circ}$
                                    $\&$ P$_t>$ 3\,GeV,
            \def\labelitemiii{--}
              \vspace{-0.1cm}
            \item for $\mu^-$\,: 20$^{\circ}$ $<\theta<$ 160$^{\circ}$
                                    $\&$ P$_t>$ 3\,GeV.
          \end{itemize}
      \end{itemize}
 \end{itemize}    
Following is the corresponding {\tt grape.cards}.
The same file is put in the directory {\tt sample}.

\vspace{0.2cm}
{\small
\renewcommand{\baselinestretch}{0.86}
\begin{verbatim}
LIST
NCALL   1400000
C ======================================================================
C << Polarization of the Lepton Beam >>
C            (1)    (2)    (3)
EPOL         -0.7    0.     0.
C ======================================================================
C << Process in the Proton Vertex >>   (1:elastic, 2:quasi-elastic, 3:DIS)
PROCESS       3
C ======================================================================
C << Produced Lepton-pair >>   (1:di-e, 2:di-mu, 3:di-tau)
LPAIR         2
C ======================================================================
C << Scattered Quark in DIS >>
C   (1:u, 2:u-bar, 3:d, 4:d-bar, 5:s, 6:s-bar, 7:c, ... , 12:t-bar)
QFLV          1
MERGE    123456
C ======================================================================
C << PDF set in DIS >>   (See PDFLIB manual.)
NGROUP        5
NSET          5
C ======================================================================
C ######################################################################
C ======================================================================
C << Electroweak Dilepton Production >>
GRASEL        2
C ======================================================================
C ######################################################################
C ======================================================================
C << Mass Range for the Hadronic System >>  (only for quasi-elastic and DIS)
MHAD          5.   300.
C ------------
C << Q2 Range for the Proton Vertex >>
Q2P           1.     1.E20
C ------------
C << Cuts for each Final-state Particle >>
C           <p/q>  <e+->  <l-+>  <l+->
THMIN        10.     0.     0.     0.
THMAX       180.   180.   180.   180.
EMIN          0.     0.     0.     0.
EMAX          1.E20  1.E20  1.E20  1.E20
PMIN          0.     0.     0.     0.
PMAX          1.E20  1.E20  1.E20  1.E20
PTMIN        15.     0.     0.     0.
PTMAX         1.E20  1.E20  1.E20  1.E20
C ------------
C << Mass cuts >>
MASSLL        4.     1.E20
              0.     1.E20
C ------------
C << Cuts for One or Some of the Final-state Particles >>
IVISI         2
C           <p/q>  <e+->  <l-+>  <l+->
THEVMIN       0.     5.    20.    20.
THEVMAX       0.   175.   160.   160.
EVMIN         0.     0.     0.     0.
EVMAX         0.     1.E20  1.E20  1.E20
PTVMIN        0.     5.     3.     3.
PTVMAX        0.     1.E20  1.E20  1.E20
C ======================================================================
STOP
\end{verbatim}
}

\clearpage
\subsection*{A part of the standard output from {\tt integ}}

{\footnotesize
\renewcommand{\baselinestretch}{0.8}
\begin{verbatim}  
      ####################################################################
      ##****************************************************************##
      ##**++++++++++++++++++++++++++++++++++++++++++++++++++++++++++++**##
      ##**++   GGGGGG     RRRRRR        A      PPPPPPP     EEEEEE   ++**##
      ##**++  G      G   R      R      A A     P      P   E         ++**##
      ##**++  G          R       R    A   A    P       P  E         ++**##
      ##**++  G  GGGGGG  R  RRRR     A     A   PPPPPPPP   EEEEEEE   ++**##
      ##**++  G     G G  R   R       AAAAAAA   P          E         ++**##
      ##**++  G     G G  R    R     A       A  P          E         ++**##
      ##**++   GGGGG  G  R      RR  A       A  P          EEEEEEEE  ++**##
      ##**++                                                        ++**##
      ##**++  GRAce-based generator for Proton-Electron collisions  ++**##
      ##**++                                                        ++**##
      ##**++               GRAPE-Dilepton_version1.1                ++**##
      ##**++                     ^^^^^^^^                           ++**##
      ##**++                      Mar. 14 2000                      ++**##
      ##**++   Comments/bug-report to Tetsuo ABE(tabe@post.kek.jp)  ++**##
      ##**++                                    (abe@mail.desy.de)  ++**##
      ##**++++++++++++++++++++++++++++++++++++++++++++++++++++++++++++**##
      ##****************************************************************##
      ####################################################################
  
                <<<<<<<<<< This is an INTEGRATION step. >>>>>>>>>>              
  
 ---> DIS process
        (Scattering of e and u quark) 
 ---> Muon-pair production 
  
                    grace 2.1(5)

          (c)Copyright 1990-1998 Minami-Tateya Group (Japan)
  
 >>> Graph selection
 jselg =
   1 :  0
   2 :  0
   3 :  0
   4 :  0
   5 :  0
   6 :  0
   7 :  0
   8 :  0
   9 :  1
   10 :  0
   11 :  0
   12 :  0
   13 :  1
   14 :  0
   15 :  0
   16 :  0
   17 :  0
   18 :  0
   19 :  0
   20 :  0
   21 :  0
   22 :  0
   23 :  0
   24 :  0
   25 :  0
  
 ========> Start of Kinematics Initialization
  
 ********** Information (in Lab. frame) **********
                               (in unit of GeV)
   P of electrons    =    27.5200
   P of protons      =    820.000
   Mass of electron  =    5.10999E-04
   Mass of proton    =   0.938272
   sqrt(S)           =    300.444
   P of CMS          =    792.480
   E of CMS          =    847.521
   gamma of CMS      =    2.82089
   beta*gamma of CMS =    2.63770
 ***********************************************
  
  << Mass range for the hadronic system >> 
       Min. =    5.00000 GeV
       Max. =    300.000 GeV
  
 ------> PDFLIB Initialization started
  
  *****  PDFLIB Version:   7.09  Released on   970702  at   16.05 
  in the CERN Computer Program Library  W5051  *****
  *****  Library compiled on  970702 at  16.05  *****

  Parm =    Nptype   Ngroup   Nset  
  Val  =    1.0000   5.0000   5.0000

  Nptype = 1  Ngroup = 5  Nset =  5  Name  = "GRV94-LO"  CrMode =  -1
  Nfl    = -5,  LO = 1,  Tmas =  180.00 GeV/c**2
  QCDL4  =  0.2000 GeV,  QCDL5 =  0.1530 GeV
  Xmin   =  0.10E-05,  Xmax = 0.99999E+00,
  Q2min  =  0.400 (GeV/c)**2,  Q2max = 0.10E+07 (GeV/c)**2
  
 ------> PDFLIB Initialization finished
  
 ------> ISR for incoming lepton  using Structure Function method 
  
 ========> End of Kinematics Initialization
  
 >>> e+ beam
  
                                                       Date:  0/ 4/12  21:29
        **********************************************************
        *                                                        *
        *     BBBBBBB     AAAA     SSSSSS   EEEEEE   SSSSSS      *
        *     BB    BB   AA  AA   SS    SS  EE      SS    SS     *
        *     BB    BB  AA    AA  SS        EE      SS           *
        *     BBBBBBB   AAAAAAAA   SSSSSS   EEEEEE   SSSSSS      *
        *     BB    BB  AA    AA        SS  EE            SS     *
        *     BB    BB  AA    AA  SS    SS  EE      SS    SS     *
        *     BBBB BB   AA    AA   SSSSSS   EEEEEE   SSSSSS      *
        *                                                        *
        *                   BASES Version 5.1                    *
        *           coded by S.Kawabata KEK, March 1994          *
        **********************************************************


     <<   Parameters for BASES    >>

      (1) Dimensions of integration etc.
          # of dimensions :    Ndim    =        9   ( 50 at max.)
          # of Wilds      :    Nwild   =        7   ( 15 at max.)
          # of sample points : Ncall   =  1399680(real)  1400000(given)
          # of subregions    : Ng      =       48 / variable
          # of regions       : Nregion =        6 / variable
          # of Hypercubes    : Ncube   =   279936

      (2) About the integration variables
          ------+---------------+---------------+-------+-------
              i       XL(i)           XU(i)       IG(i)   Wild
          ------+---------------+---------------+-------+-------
              1    0.000000E+00    1.000000E+00     1      yes
              2    0.000000E+00    1.000000E+00     1      yes
              3    0.000000E+00    1.000000E+00     1      yes
              4    0.000000E+00    1.000000E+00     1      yes
              5    0.000000E+00    1.000000E+00     1      yes
              6    0.000000E+00    1.000000E+00     1      yes
              7    0.000000E+00    1.000000E+00     1      yes
              8    0.000000E+00    1.000000E+00     0       no
              9    0.000000E+00    1.000000E+00     0       no
          ------+---------------+---------------+-------+-------

      (3) Parameters for the grid optimization step
          Max.# of iterations: ITMX1 =        4
          Expected accuracy  : Acc1  =   0.2000 %

      (4) Parameters for the integration step
          Max.# of iterations: ITMX2 =       10
          Expected accuracy  : Acc2  =   0.0100 %

                                                       Date:  0/ 4/12  21:29
               Convergency Behavior for the Grid Optimization Step
 ------------------------------------------------------------------------------
 <- Result of  each iteration ->  <-     Cumulative Result     -> < CPU  time >
  IT Eff R_Neg   Estimate  Acc %  Estimate(+- Error )order  Acc % ( H: M: Sec )
 ------------------------------------------------------------------------------
   1   2  0.00  4.586E-02  6.435  4.586406(+-0.295122)E-02  6.435   0: 1:37.57
   2  29  0.00  5.205E-02  0.295  5.203249(+-0.015359)E-02  0.295   0: 7:23.86
   3  37  0.00  5.196E-02  0.201  5.198557(+-0.008625)E-02  0.166   0:13:52.62
 ------------------------------------------------------------------------------

                                                       Date:  0/ 4/12  21:29
               Convergency Behavior for the Integration Step      
 ------------------------------------------------------------------------------
 <- Result of  each iteration ->  <-     Cumulative Result     -> < CPU  time >
  IT Eff R_Neg   Estimate  Acc %  Estimate(+- Error )order  Acc % ( H: M: Sec )
 ------------------------------------------------------------------------------
   1  38  0.00  5.196E-02  0.194  5.196162(+-0.010077)E-02  0.194   0:20:27.03
   2  39  0.00  5.201E-02  0.198  5.198416(+-0.007197)E-02  0.138   0:27: 0.89
   3  38  0.00  5.201E-02  0.395  5.198705(+-0.006793)E-02  0.131   0:33:34.96
   4  39  0.00  5.199E-02  0.191  5.198759(+-0.005605)E-02  0.108   0:40: 9.71
   5  38  0.00  5.200E-02  0.195  5.198989(+-0.004905)E-02  0.094   0:46:43.94
   6  39  0.00  5.200E-02  0.193  5.199218(+-0.004405)E-02  0.085   0:53:18.60
   7  38  0.00  5.187E-02  0.193  5.197301(+-0.004032)E-02  0.078   0:59:53.16
   8  38  0.00  5.185E-02  0.193  5.195602(+-0.003739)E-02  0.072   1: 6:27.22
   9  38  0.00  5.205E-02  0.193  5.196769(+-0.003505)E-02  0.067   1:13: 1.21
  10  38  0.00  5.191E-02  0.196  5.196186(+-0.003313)E-02  0.064   1:19:34.81
 ------------------------------------------------------------------------------

                    ****** END OF BASES *********


     <<   Computing Time Information   >>

               (1) For BASES              H: M:  Sec
                   Overhead           :   0: 0: 0.12
                   Grid Optim. Step   :   0:13:52.62
                   Integration Step   :   1: 5:42.20
                   Go time for all    :   1:19:34.94

               (2) Expected event generation time
                   Expected time for 1000 events :      0.40 Sec
  
 Making bases.rz...
    ---> BASE1 : finished
    ---> BASE3 : finished
    ---> BASE4 : finished
    ---> BASE5 : finished
    ---> RANDM : finished
    ---> PLOTH : finished
    ---> PLOTB : finished
    ---> BSRSLT: finished

 ===> Directory : //bn
          3 (N)   BASES data(real*8)  
          1 (N)   BASES data(integer*4)   
          2 (N)   BASES data(real*4)  
\end{verbatim}
}

\clearpage
\subsection*{A part of the standard output from {\tt spring}}

{\footnotesize
\renewcommand{\baselinestretch}{0.8}
\begin{verbatim}
 Suppressing decay of the following particles in PYTHIA,
   K_S0       eta        Lambda0    Sigma+     Sigma0     Sigma-     Xi0       
   Xi-        Omega-     D+         D0         D_s+       Lambda_c+  mu-       
   tau-       pi+        K+         K_L0       pi0       
  
 Loading bases.rz...
    ---> BASE1 : finished
    ---> BASE3 : finished
    ---> BASE4 : finished
    ---> BASE5 : finished
    ---> RANDM : finished
    ---> PLOTH : finished
    ---> PLOTB : finished
    ---> BSRSLT: finished
  
1                                                                              
 ******************************************************************************
 ******************************************************************************
 **                                                                          **
 **                                                                          **
 **              *......*                  Welcome to the Lund Monte Carlo!  **
 **         *:::!!:::::::::::*                                               **
 **      *::::::!!::::::::::::::*          PPP  Y   Y TTTTT H   H III   A    **
 **    *::::::::!!::::::::::::::::*        P  P  Y Y    T   H   H  I   A A   **
 **   *:::::::::!!:::::::::::::::::*       PPP    Y     T   HHHHH  I  AAAAA  **
 **   *:::::::::!!:::::::::::::::::*       P      Y     T   H   H  I  A   A  **
 **    *::::::::!!::::::::::::::::*!       P      Y     T   H   H III A   A  **
 **      *::::::!!::::::::::::::* !!                                         **
 **      !! *:::!!:::::::::::*    !!       This is PYTHIA version 6.136      **
 **      !!     !* -><- *         !!       Last date of change: 30 Nov 1999  **
 **      !!     !!                !!                                         **
 **      !!     !!                !!                                         **
 **      !!                       !!                                         **
 **      !!        ep             !!       Disclaimer: this program comes    **
 **      !!                       !!       without any guarantees. Beware    **
 **      !!                 pp    !!       of errors and use common sense    **
 **      !!   e+e-                !!       when interpreting results.        **
 **      !!                       !!                                         **
 **      !!                                Copyright T. Sjostrand (1999)     **
 **                                                                          **
 ** An archive of program versions and documentation is found on the web:    **
 ** http://www.thep.lu.se/~torbjorn/Pythia.html                              **
 **                                                                          **
 ** When you cite this program, currently the official reference is          **
 ** T. Sjostrand, Computer Physics Commun. 82 (1994) 74.                     **
 ** The supersymmetry extensions are described in                            **
 ** S. Mrenna, Computer Physics Commun. 101 (1997) 232                       **
 ** Also remember that the program, to a large extent, represents original   **
 ** physics research. Other publications of special relevance to your        **
 ** studies may therefore deserve separate mention.                          **
 **                                                                          **
 ** Main author: Torbjorn Sjostrand; Department of Theoretical Physics 2,    **
 **   Lund University, Solvegatan 14A, S-223 62 Lund, Sweden;                **
 **   phone: + 46 - 46 - 222 48 16; e-mail: torbjorn@thep.lu.se              **
 ** SUSY author: Stephen Mrenna, Physics Department, UC Davis,               **
 **   One Shields Avenue, Davis, CA 95616, USA;                              **
 **   phone: + 1 - 530 - 752 - 2661; e-mail: mrenna@physics.ucdavis.edu      **
 **                                                                          **
 **                                                                          **
 ******************************************************************************
 ******************************************************************************
1****************** PYINIT: initialization of PYTHIA routines *****************

 ==============================================================================
 I                                                                            I
 I         PYTHIA will be initialized for p+ on e+ user configuration         I
 I                                                                            I
 I                  px (GeV/c)   py (GeV/c)   pz (GeV/c)      E (GeV)         I
 I        p+             0.000        0.000      820.000      820.001         I
 I        e+             0.000        0.000      -27.520       27.520         I
 I                                                                            I
 I           corresponding to    300.444 GeV center-of-mass energy            I
 I                                                                            I
 ==============================================================================

 ******** PYMAXI: summary of differential cross-section maximum search ********

           ==========================================================
           I                                      I                 I
           I  ISUB  Subprocess name               I  Maximum value  I
           I                                      I                 I
           ==========================================================
           I                                      I                 I
           I  308   e+ uu^dd^ss^ -> e+ q m+ m-    I    5.1962D-02   I
           I                                      I                 I
           ==========================================================

 ********************** PYINIT: initialization completed **********************


==========> START of SPRING at  1:21(13/ 4/ 0)
  
 Number of generated events =  100
  
                            Event listing (summary)

    I particle/jet KS     KF  orig    p_x      p_y      p_z       E        m

    1 !p+!         21    2212    0    0.000    0.000  820.000  820.001    0.938
    2 !e+!         21     -11    0    0.000    0.000  -27.520   27.520    0.001
 ==============================================================================
    3 !d!          21       1    1   -0.343   -0.058  168.409  168.409    0.000
    4 !e+!         21     -11    2    0.000    0.000  -27.520   27.520    0.000
    5 !d!          21       1    3    1.167   -0.401  138.481  138.487    0.000
    6 !e+!         21     -11    4    0.000    0.000  -27.497   27.497    0.000
    7 !d!          21       1    0  -10.149   16.228   96.119   98.007    0.004
    8 !e+!         21     -11    0    0.009    0.020   -6.958    6.958    0.001
    9 !mu-!        21      13    0   19.804   -7.523  -12.433   24.564    0.106
   10 !mu+!        21     -13    0   -8.498   -9.126   34.256   36.455    0.106
 ==============================================================================
   11 e+            1     -11    8    0.009    0.020   -6.958    6.958    0.001
   12 mu-           1      13    9   19.804   -7.523  -12.433   24.564    0.106
   13 gamma         1      22    4    0.000    0.000    0.000    0.000    0.000
   14 mu+           1     -13   10   -7.882   -8.464   31.770   33.810    0.106
   15 gamma         1      22    2    0.000    0.000    0.000    0.000    0.000
   16 (d)       A  12       1    7   -9.379   15.009   77.561   79.555    0.004
   17 (g)       I  12      21    7   -1.386    0.556   21.044   21.097    0.000
   18 (g)       I  12      21    3   -1.510    0.343   29.905   29.945    0.000
   19 (uu_1)    V  11    2203    1    0.343    0.058  651.591  651.591    0.771
 ==============================================================================
   20 (string)     11      92   16  -11.932   15.967  780.101  782.188   53.513
   21 (rho-)       11    -213   20   -2.614    3.803   19.032   19.599    0.763
   22 eta           1     221   20   -5.726    8.838   49.702   50.808    0.547
   23 (K*+)        11     323   20   -0.313    0.722    5.628    5.751    0.885
   24 (K*-)        11    -323   20   -1.601    1.458   13.870   14.064    0.862
   25 (rho+)       11     213   20   -0.642    0.548   18.730   18.778    1.033
   26 pi0           1     111   20   -0.205    0.308    3.644    3.665    0.135
   27 (rho-)       11    -213   20   -0.597   -0.035   42.069   42.081    0.796
   28 (rho0)       11     113   20   -0.171    0.378   46.359   46.369    0.877
   29 pi0           1     111   20    0.093   -0.222   22.348   22.349    0.135
   30 K+            1     321   20   -0.371    0.069   95.962   95.964    0.494
   31 Sigma0        1    3212   20    0.358    0.006  317.588  317.590    1.193
   32 pi+           1     211   20   -0.143    0.094  145.170  145.170    0.140
   33 pi-           1    -211   21   -1.237    1.578    6.910    7.196    0.140
   34 pi0           1     111   21   -1.376    2.225   12.122   12.402    0.135
   35 K+            1     321   23   -0.236    0.131    2.313    2.380    0.494
   36 pi0           1     111   23   -0.078    0.591    3.315    3.370    0.135
   37 (Kbar0)      11    -311   24   -0.822    0.679    8.481    8.562    0.498
   38 pi-           1    -211   24   -0.779    0.779    5.389    5.502    0.140
   39 pi+           1     211   25   -0.837    0.227   15.119   15.144    0.140
   40 pi0           1     111   25    0.195    0.321    3.611    3.633    0.135
   41 pi-           1    -211   27   -0.260    0.270   29.720   29.722    0.140
   42 pi0           1     111   27   -0.337   -0.305   12.350   12.359    0.135
   43 pi+           1     211   28   -0.149   -0.213    8.213    8.218    0.140
   44 pi-           1    -211   28   -0.022    0.591   38.146   38.151    0.140
   45 K_S0          1     310   37   -0.822    0.679    8.481    8.562    0.498
 ==============================================================================
                   sum:  2.00         0.000    0.000  792.480  847.521  300.444


 ===> Directory : //grp
==========> END   of SPRING at  1:21(13/ 4/ 0)
  
1********* PYSTAT:  Statistics on Number of Events and Cross-sections *********

 ==============================================================================
 I                                  I                            I            I
 I            Subprocess            I      Number of points      I    Sigma   I
 I                                  I                            I            I
 I----------------------------------I----------------------------I    (pb)    I
 I                                  I                            I            I
 I N:o Type                         I    Generated         Tried I            I
 I                                  I                            I            I
 ==============================================================================
 I                                  I                            I            I
 I   0 All included subprocesses    I          100           100 I  5.196D-02 I
 I 308 e+ uu^dd^ss^ -> e+ q m+ m-   I          100           100 I  5.196D-02 I
 I                                  I                            I            I
 ==============================================================================

 ********* Fraction of events that fail fragmentation cuts =  0.00000 *********

             ************* << Total Cross-section >> *************
             *                                                   *
             *          ( 5.196186 +- 0.003313 )E-02 pb          *
             *                                                   *
             *****************************************************
\end{verbatim}
}

\end{document}